\documentstyle[12pt]{article}
\def\bbox#1{{\bf #1}}
\def\text#1{{\rm #1}}
\def\Journal#1#2#3#4{{#1} {\bf #2}, #3 (#4)}
\def\Book#1#2#3{{\em #1} ({#2}, {#3})}

\begin{document}
\title{M\"obius invariant integrable lattice equations
associated with KP and 2DTL hierarchies}
\author{L. V. Bogdanov\thanks{
Permanent address: L. D. Landau Institute for Theoretical Physics, Kosygin
str. 2, Moscow 117940, GSP1, Russia; e-mail Leonid@landau.ac.ru}
\hspace{0.01em} and B. G. Konopelchenko
\thanks{
Also: Budker Institute of Nuclear Physics, Novosibirsk 90, Russia}\\
\small Consortium EINSTEIN\thanks{
European Institute for Nonlinear Studies via Transnationally Extended
Interchanges}, \\
\small Dipartimento di Fisica dell'Universit\`a di Lecce \\
\small and Sezione INFN,
73100 Lecce, Italy}
\date{}
\maketitle
\begin{abstract}
Integrable lattice equations arising in the context
of singular manifold equations
for scalar, multicomponent KP
hierarchies and 2D Toda lattice hierarchy are considered.
These equation generate the corresponding continuous hierarchy
of singular manifold equations,
its B\"acklund transformations and different forms of
superposition principles. They possess rather special form
of compatibility representation. The distinctive feature
of these equations is invariance under the action of
M\"obius transformation. Geometric interpretation
of these discrete equations is given.
\end{abstract}
\subsection*{Introduction}
The equations we are going to discuss here were
derived by the authors in frame of analytic-bilinear
approach to integrable hierarchies \cite{AB1},
\cite{AB2}. They arise on the third level
of the hierarchy, i.e. as a discrete
singular manifold equations (or, in other
words, in the context  of
the hierarchy in Schwarzian form). These
equations are highly symmetric and possess very
peculiar properties.

First, they have a very special compatibility
condition representation, symmetric with respect
to lattice variables and indicating duality
between wave functions and potential
(providing in fact equations for both objects
in a similar way).

Second, they possess (even in the matrix case,
corresponding to multicomponent KP hierarchy)
a M\"obius symmetry, which is typical for
equations in Schwarzian form and which is
deeply connected to projective geometry.
Similar equations for the KdV case were interpreted
by Bobenko and Pinkall \cite{Bobenko} as equations of discrete
isothermic surfaces.
We hope that equations we discuss may also
be interpreted in terms of some special classes
of discrete surfaces, though the interpretation
is yet unknown for the matrix case. In the scalar
case, however, there are at least two ways to connect
the equations to discrete geometry: as a reduction
of Darboux system and in terms of the three-dimensional
lattice on the complex plane with some special
geometrical properties.

Third, for the basic hierarchy we have
{\em  one} lattice equation of the considered type,
and it generates by expansion in parameters
the continuous hierarchy itself, its
B\"acklund transformations and different types
of superposition formulae for B\"acklund transformations.
Thus it encodes all the information about the
hierarchy (more or less in the way of Hirota
bilinear discrete equation, with which it is
closely connected, but not in terms of the
$\tau$-function, rather in terms of
wave functions and potentials).
\subsection*{Basic equations and their properties}
In this section we introduce basic equations and
study some of their properties without the reference
to the original scheme of derivation of these equations
(a sketch of it will be given later).

In the context
of the scalar KP hierarchy the basic equation reads
\begin{equation}
(T_2\Delta_1 \Phi)(T_3\Delta_2 \Phi)(T_1\Delta_3 \Phi)=
(T_2\Delta_3 \Phi)(T_3\Delta_1 \Phi)(T_1\Delta_2 \Phi).
\label{1KPSM}
\end{equation}
Here
$$
\Phi=\Phi(n_1,n_2,n_3),
$$
$$
T_1\Phi=\Phi(n_1+1,n_2,n_3),
$$
$$
\Delta_1=T_1-1.
$$
In the case of multicomponent KP hierarchy $\Phi$
is $N\times N$ matrix-valued function and the basic
equation looks like
\begin{equation}
(T_1\Delta_3 \Phi)\cdot(T_3\Delta_1 \Phi)^{-1}\cdot
(T_3\Delta_2 \Phi)\cdot(T_2\Delta_3 \Phi)^{-1}\cdot
(T_2\Delta_1 \Phi)\cdot(T_1\Delta_2 \Phi)^{-1}=1.
\label{KPsingman}
\end{equation}
And finally, for the scalar 2D Toda lattice hierarchy
one has
\begin{eqnarray}
&&
\{T_-(T_+-1)\Phi\}\{T(T_--T^{-1})\Phi\}\{T_+(1-T^{-1})\Phi\}=
\nonumber\\&&
\{(T_+-1)\Phi\}\{T_+(T_--T^{-1})\Phi\}\{T_-(T-1)\Phi\}.
\label{TodaSM}
\end{eqnarray}
This equation is not symmetric with respect to
all three shifts, one of them ($T$) plays a special role
and corresponds to the original (essentially) discrete
variable of the Toda lattice.
In what follows we will mainly study
equations (\ref{1KPSM}), (\ref{KPsingman})

Equation (\ref{1KPSM}) may be treated
as compatibility condition
for the following set of equations
for the function $f$
\begin{equation}
{a_i T_i f\over a_j T_j f}=
{\Delta_i\Phi\over \Delta_j \Phi}
\quad i,j,k\in \{1,2,3\},
\label{1KPSMlinear}
\end{equation}
where $a_i$ are some constants. To demonstrate this,
we break the symmetry of the relation (\ref{1KPSMlinear})
and distinguish one of the shifts (say, $T_1$).\\
{\bf Remark.}
{\small Unfortunately, we don't know a symmetric derivation
starting from relation (\ref{1KPSMlinear}),
though the initial equation and the final equation
are symmetric. Symmetric way of deriving (\ref{1KPSM})
in another context will be given later in this paper.}
Then we rewrite (\ref{1KPSMlinear}) in the form
$$
T_2 f= {a_1\Delta_2 \Phi\over a_2\Delta_1\Phi}
T_1 f=U T_1f,
$$
$$
T_3 f= {a_1\Delta_3 \Phi\over a_3\Delta_1\Phi}
T_1 f=V T_1f.
$$
In this form the system looks more familiar.
Then, as usual, taking cross-shifts of the first and the second
equations (which should give the same results) and using the
equations themselves to get rid of the shifts $T_2$, $T_3$
acting on $f$, one obtains the following equation
(compatibility condition)
$$
(T_3 U)(T_1 V)=(T_2 V)(T_1 U).
$$
Substituting the expressions of $U$, $V$ through
$\Phi$, one gets exactly equation
(\ref{1KPSM}).

It is possible to start also with the following
(`dual') system to get equation (\ref{1KPSM})
\begin{equation}
{a_i T_i^{-1} \widetilde f
\over a_j T_j^{-1} \widetilde f}=
{\widetilde\Delta_i\Phi\over \widetilde\Delta_j \Phi}
\quad i,j,k\in \{1,2,3\},
\label{d1KPSMlinear}
\end{equation}
where
$$
\widetilde \Delta_i=T_i^{-1}-1.
$$

On the other hand, it is possible to treat
the system (\ref{1KPSMlinear}) as a set of linear
equations for the function $\Phi$. To do it,
we rewrite the system in the form
$$
\Delta_2\Phi={a_2 T_2 f\over a_1 T_1 f}\Delta_1 f=
U\Delta_1 \Phi,
$$
$$
\Delta_3\Phi={a_3 T_3 f\over a_1 T_1 f}\Delta_1 f=
V\Delta_1 \Phi.
$$
The compatibility condition for this system gives
two equations
$$
(T_3 U)(T_1 V)=(T_2V)(T_1 U),
$$
$$
\Delta_3 U+(T_3U)\Delta_1V=
\Delta _2V+ (T_2V)\Delta_1U.
$$
The first of these equations is resolved by the substitution
of expressions for the functions $U$, $V$ in terms of
$f$, the second gives the following equation for~$f$
\begin{equation}
\sum_{(ijk)}\epsilon_{ijk}a_ja_kT_j
\left({f\over T_i f}
\right)=0,
\label{dKPwave}
\end{equation}
summation here is over different permutations of
indices.

Starting from the dual linear system
(\ref{d1KPSMlinear}), one obtains an equation for the
function $\widetilde f$
\begin{equation}
\sum_{(ijk)}\epsilon_{ijk}a_ja_kT_k
\left({T_i\widetilde f\over \widetilde f}
\right)=0,
\label{KPwave}
\end{equation}
Both these equation are connected with the modified
KP hierarchy (the
equation for $\widetilde f$ with the mKP hierarchy
itself and the equation for $f$ with the dual
(adjoint) hierarchy).
Similar equations  were derived by Nijhoff et al
\cite{Nijhoff}.

The matrix case (\ref{KPsingman}) can be treated
the same way. Equations (\ref{d1KPSMlinear})
and the dual system
for this case read
$$
(\Delta_j \Phi)^{-1}\Delta_i \Phi=
(A_j T_j f)^{-1}(A_i T_i f),
$$
$$
\widetilde\Delta_i \Phi(\widetilde\Delta_j \Phi)^{-1}=
(T_i^{-1}\widetilde f A_i)(T_j^{-1} \widetilde f A_j)^{-1},
$$
where $A_i$ are some diagonal matrices;
in what follows we suggest that the determinant
of $A_1$ is not equal to zero.

The linear system for the function $f$ takes the form
$$
A_1^{-1}A_2T_2 f=(T_1 f)(\Delta_1 \Phi)^{-1}\Delta_2\Phi
=(T_1 f)U,
$$
$$
A_1^{-1}A_3T_3 f=(T_1 f)(\Delta_1 \Phi)^{-1}\Delta_3\Phi
=(T_1 f)V.
$$
The compatibility condition for this system
gives equation (\ref{KPsingman}) in the form
$$
(T_1\Delta_3 \Phi)\cdot(T_3\Delta_1 \Phi)^{-1}\cdot
(T_3\Delta_2 \Phi)=(T_1\Delta_2 \Phi)\cdot
(T_2\Delta_1 \Phi)^{-1}\cdot(T_2\Delta_3 \Phi).
$$
Then, the linear system for the function
$\Phi$ reads
$$
\Delta_2\Phi=\Delta_1\Phi(a_1T_1f)^{-1}A_2T_2 f=
\Delta_1\Phi U,
$$
$$
\Delta_3\Phi=\Delta_1\Phi(a_1T_1f)^{-1}A_3T_3 f=
\Delta_1\Phi V,
$$
and a compatibility condition gives a matrix version
of equation (\ref{dKPwave})
$$
\sum_{(ijk)}\epsilon_{ijk}A_jT_j\left(f
\cdot (T_if )^{-1}\right)A_k=0.
$$
A matrix version of (\ref{KPwave}) is
$$
\sum_{(ijk)}\epsilon_{ijk}A_kT_k\left(\widetilde f^{-1}
\cdot(T_i\widetilde f)
\right)A_j=0.
$$
\subsection*{Symmetries of the basic equations}
It is easy to check that equation (\ref{1KPSM}), (\ref{TodaSM})
possess a M\"obius symmetry transformation,
i.e. if function $\Phi$ is a solution
to this equation, then  function $\Phi'$,
$$
\Phi'={a\Phi+b\over c\Phi+d},
$$
where $a,b,c,d$ are arbitrary constants,
is also a solution. For the matrix equation
(\ref{KPsingman}) the symmetries include inversion,
left and right matrix multiplication and shift
$$
\Phi'=\Phi^{-1},
$$
$$
\Phi'=A\Phi,\quad \Phi'=\Phi B,
$$
$$
\Phi'=\Phi+C,
$$
where $A,B,C$ are some matrices (determinant
of $A$ and $B$ is not equal to zero).

The structure
of equation (\ref{KPsingman}) resembles the
structure of quartic equations
given by Bobenko
and Pinkall \cite{Bobenko} for the quaternion description
of discrete isothermic surfaces in ${\bf R^3}$.
Probably equation (\ref{KPsingman})  also
has a geometric interpretation in terms of
quaternion construction.

It is interesting to note that due to the inversion
symmetry, the equation (\ref{KPsingman})
admits a reduction
$$
\Phi\Phi=-1,
$$
that for $\Phi$ belonging to the field of quaternions
means that the vector in ${\bf R^3}$ corresponding
to the (imaginary) quaternion $\Phi$ has a unit
length, and so the equation (\ref{KPsingman})
in this case defines a three-dimensional lattice
on the unit sphere.

Equations (\ref{dKPwave}), (\ref{KPwave}) are
connected by the transformation
$$
f'={\widetilde f}^{-1}\,,
$$
$$
\widetilde f'={f}^{-1}\,,
$$
that keeps in the matrix case.
\subsection*{The hierarchy of continuous equations}
Let us suggest that the function $\Phi$ in equation
(\ref{1KPSM}) is a function of a standard infinite
set of KP variables ${\bf x}$,
${\bf x}=\{x_k\},\quad 1\leq k<\infty$ (we will also
use notations $x=x_1$, $y=x_2$, $t=x_3$) and
the shift operators $T_i$ are realized as
\begin{equation}
T_i:{\bf x}\rightarrow{\bf x}+[a_i],
\label{shifts}
\end{equation}
$$
[a_i]=\{{1\over k}a_i^k\}.
$$
This suggestion is justified by the explicit
construction given in \cite{AB1}, \cite{AB2}.
Then, expanding (\ref{1KPSM}) into the powers
of $a_1$ and taking the first term, we get
\begin{equation}
(T_2 \Phi_x)(T_3\Delta_2\Phi)\Delta_3\Phi=
(T_3\Phi_x)(T_2\Delta_3\Phi)\Delta_2\Phi.
\label{SM2-1}
\end{equation}
The next step is to expand this equation
in $a_2$. The first nonvanishing term
arises at the second order in $a_2$.
After simple transformations we get
\begin{equation}
{\partial\over \partial x}
\ln \left({1\over \Phi_x}
{\Delta \Phi\over a}\right)=
{1\over 2}\Delta\left({\Phi_y+\Phi_{xx}\over
\Phi_x}\right)
\label{SM1-2},
\end{equation}
where we are left with one discrete variable
and parameter $a$, so we omit the index 3.
And finally, expanding in parameter and taking
the first nonvanishing term, which corresponds to the
third order in $a_3$ in the initial equation
(\ref{1KPSM}), we obtain
\begin{eqnarray}
&&\Phi_t=\mbox{${1\over4}$}\Phi_{xxx}+\mbox{${3\over8}$} {\frac{
\Phi_y^2-\Phi_{xx}^2}{\Phi_x}}+ \mbox{${3\over4}$}\Phi_x W_y ,
\quad W_x={
\frac{\Phi_y}{\Phi_x}}\,.
\label{singman}
\end{eqnarray}
This equation first arose in Painleve analysis of the KP
equation as a
singular manifold equation \cite{Weiss}.
The higher terms of expansion of equation
(\ref{1KPSM}) will lead to the hierarchy of singular
manifold equations.

So we have the following objects:\\
1. Lattice equation (\ref{1KPSM})\\
2. Equation with two discrete and one continuous
variables (\ref{SM2-1})\\
3. Equation with one discrete and
two continuous variables (\ref{SM1-2})\\
4. PDE (\ref{singman}) with three continuous variables \\
(we would like to emphasize that for the cases
2, 3, 4 we have an infinite hierarchy of equations).
All these equations are the symmetries of each
other by construction, and can be interpreted
in different ways (continuous symmetries of discrete
equations, discrete symmetries of continuous equations
and mixed cases). The interpretation depends on the
choice of the basic equation (i.e. in some
sense on the point of reference).

A standard way is to take continuous equation
(\ref{singman})
as a basic system. Then the interpretation
of the other objects is the following:\\
3. Equation (\ref{SM1-2}) defines a B\"acklund
transformation for the equation (\ref{singman})\\
2. Equation (\ref{SM2-1}) is a superposition
principle for two B\"acklund transformations\\
1. Lattice equation (\ref{1KPSM})
provides an {\em algebraic} superposition principle
for three B\"acklund transformations.
\subsection*{Derivation of equations (\ref{1KPSM}),
(\ref{KPsingman}) from the equations for the CBA
function}
In \cite{AB1}, \cite{AB2} the following equations
were derived for the scalar KP
Cauchy-Baker-Akhiezer (CBA) function
\begin{equation}
\Delta_i\Psi(\lambda,\mu,{\bf x})=
a_i\widetilde\psi(\mu,{\bf x})T_i\psi(\lambda,{\bf x}),
\label{CBA}
\end{equation}
where $\Psi(\lambda,\mu,{\bf x})$ is a CBA function,
$$
\Psi(\lambda,\mu,{\bf x})=g(\lambda,\bbox{x})
\chi(\lambda,\mu,\bbox{x})g^{-1}(\mu,\bbox{x}),
$$
$$
g(\lambda,\bbox{x})=\exp \left(\sum_{n=1}^{\infty}
x_n \lambda^{-n}\right),
$$
$\chi(\lambda,\mu, \bbox{x})$ is defined in the
unit disc in
$\lambda,\mu$, is analytic in these variables for
$\lambda\neq\mu$ and for $\lambda\rightarrow\mu$
behaves like $(\lambda-\mu)^{-1}$;
$\psi(\lambda,{\bf x})$ and $\widetilde\psi(\mu,{\bf x})$
are respectively Baker-Akhiezer (BA) and dual (adjoint)
BA functions
$$
\psi(\lambda,{\bf x})=g(\lambda,\bbox{x})
\chi(\lambda,0,\bbox{x})
$$
$$
\widetilde\psi(\mu,{\bf x})=
\chi(0,\mu,\bbox{x})g^{-1}(\mu,\bbox{x}),
$$
the shifts $T_i$ are realized in the form
(\ref{shifts}).

For the multicomponent KP case
CBA and BA functions are $N\times N$ matrix-valued
functions, we have $N$ infinite sets of KP
variables
$$
\bbox{x}=(\bbox{x}_1,\dots,\bbox{x}_N),
$$
and the action of three shift operators
$T_i$ is defined as
$$
T_i:\bbox{x}\rightarrow \bbox{x}+[\bbox{a}_i],
$$
$$
[\bbox{a}_i]=([a_{i(1)}],\dots, [a_{i(N)}]),
$$
$\bbox{a}_i$ are some constant $N$-dimensional
vectors.
The equation (\ref{CBA}) for the multicomponent case reads
\begin{equation}
\Delta_i\Psi(\lambda,\mu,{\bf x})=
\widetilde\psi(\mu,{\bf x})A_iT_i\psi(\lambda,{\bf x}),
\label{CBAmulti}
\end{equation}
$$
A_i=\text{diag}(a_{i(1)},\dots, a_{i(N)}).
$$

Integrating the equation (\ref{CBA}) with two
arbitrary functions $\rho(\lambda)$,
$\widetilde\rho(\mu)$ over the unit circle
in $\lambda$, $\mu$, we get
\begin{equation}
\Delta_i\Phi({\bf x})=a_i\widetilde f({\bf x}) T_i f({\bf x}).
\label{CBAx}
\end{equation}
Using this equation, it is easy to check
that $\Phi$ satisfies the equation (\ref{1KPSM}).
And, taking cross-differences of equations
(\ref{CBAx}) and combining them, one gets
equations (\ref{dKPwave}), (\ref{KPwave})
in a symmetric algebraic manner, without
reference to compatibility conditions.
The multicomponent case (\ref{KPsingman})
is analogous.

An alternative way to derive equations
for $\Phi$, $f$, $\widetilde f$
from (\ref{CBAx}) is just to
reproduce equations (\ref{1KPSMlinear}), (\ref{d1KPSMlinear})
(which is straightforward), and then use
compatibility conditions.

Taking  the weight functions on the unit circle
in the form $\rho(\lambda)=\delta(\lambda_0-\lambda)$,
$\widetilde\rho(\mu)=\delta(\mu_0-\mu)$,
one arrives to the conclusion that the CBA function itself
satisfies the equation (\ref{1KPSM}) (or
(\ref{KPsingman}) in the matrix case). Then,
using the analytic properties of the CBA function
and going to the limit $\lambda\rightarrow\mu$,
one obtains for the function
$u({\bf x})=\Psi_{\rm r}(0,0,{\bf x})$ the discrete
version of the KP hierarchy in the form
of algebraic superposition principle for
three B\"acklund transformations
\begin{equation}
\sum_{(ijk)} \epsilon_{ijk}A_kT_k (\Delta_i u -uA_iT_iu)A_j=0,
\label{multiKP}
\end{equation}
where
summation goes over different permutations of
indices.
Similar equation
can be found in \cite{Nijhoff}.

It is also possible to reproduce in this way
equations (\ref{dKPwave}), (\ref{KPwave}) taking
only one weight function as $\delta$-function
and going to the limit $\lambda\rightarrow 0$
or $\mu\rightarrow 0$.

Thus the equations (\ref{1KPSM}), (\ref{KPsingman})
encode all the three levels of generalized KP
hierarchy: hierarchy of singular manifold equations,
modified hierarchy and the basic KP hierarchy.
\subsection*{Equation (\ref{1KPSM}) and quadrilateral
lattices}
A more general version of equation (\ref{CBA})
reads (we need only scalar case here) \cite{AB1}
\begin{equation}
\Psi(\lambda,\mu,{\bf x}+[a])
-\Psi(\lambda,\mu,{\bf x}+[b])=
(a-b)\widetilde\psi_b(\mu,{\bf x}+[b])\psi_b(\lambda,{\bf x}+[a]),
\label{CBAgen}
\end{equation}
where $\psi_b(\lambda,{\bf x})$, $\widetilde\psi_b(\mu,{\bf x})$
are respectively BA and dual BA function associated with the point
$b$ rather then the point zero for equation (\ref{CBA}),
$$\psi_b(\lambda,{\bf x})=
\chi(\lambda,b,\bbox{x})g(\lambda,\bbox{x}),
$$
$$
\widetilde\psi_b(\mu,{\bf x})=g^{-1}(\mu,\bbox{x})
\chi(b,\mu,\bbox{x}).
$$
For $a=0$ the equation (\ref{CBAgen})
gives
\begin{equation}
\Psi(\lambda,\mu,{\bf x}+[b])-\Psi(\lambda,\mu,{\bf x})=
b\widetilde\psi_b(\mu,{\bf x}+[b])\psi_b(\lambda,{\bf x}).
\label{CBAgen1}
\end{equation}
Introducing
lattice variables and taking $b$ equal to $a_i$,
we get
\begin{equation}
\Delta_i\Psi(\lambda,\mu,{\bf n})=
a_i\psi_{a_i}(\lambda,{\bf n})
T_i\widetilde\psi_{a_i}(\mu,{\bf n}).
\label{CBAgen2}
\end{equation}
Then, recalling analytic properties of BA and CBA
functions,  removing the common factors
$g_j(\lambda,n_j)$, $g_k(\mu,n_k)$, $k\neq i\neq j\neq k$  from
the equation (\ref{CBAgen2}),
$$
g_i(\lambda,n_i)=g(\lambda,n_i[a_i])=
\left({\lambda\over \lambda-a_i}\right)^{n_i},
$$
and taking it at the points
$\lambda=a_j$, $\mu=a_k$, we get a discrete Darboux
equation in terms of rotation coefficients
\cite{BK1}
\begin{equation}
\Delta_i \beta_{jk}=\beta_{ji}T_i\beta_{ik}
\label{dDarboux}
\end{equation}
where
$$\beta_{ij}=
\sqrt{a_ia_j}{g_j(a_i)g_k(a_i)\over g_i(a_j)g_k(a_j)}
\chi(a_i,a_j)
$$
As it was recently discovered \cite{Santini},  the equation
(\ref{dDarboux}) describes a system of planar quadrilateral
lattices.

We have derived the equation (\ref{dDarboux}) in some
very special setting, i.e. in the context of the
scalar KP hierarchy, so we could expect to have
some additional constrains. Indeed, doing some algebra,
from (\ref{CBAgen}) one also gets the following relations
\begin{eqnarray}
&&\phi_{ij}\phi_{ji}=1,
\nonumber\\
&&\phi_{ij}\phi_{jk}\phi_{ki}=\phi_{ik}\phi_{kj}\phi_{ji}=1,
\nonumber\\
&&\phi_{ij}=T_i\beta_{ij}.
\label{constraints}
\end{eqnarray}
These constraints from the first sight look a bit
mysterious, and one wonders how it is possible
to satisfy them. But recalling the expression
for the $\chi(a_i,a_j,\bbox{n})$ in terms of the $\tau$-function
\cite{AB1}
$$
\chi(a_i,a_j,\bbox{n})={1\over a_i-a_j}
{T_jT_i^{-1}\tau(\bbox{n})\over\tau(\bbox{n})},
$$
for $\phi_{ij}$ one gets
$$
\phi_{ij}=\sqrt{{a_i\over a_j}}
{g_j(a_i)g_k(a_i)\over g_i(a_j)g_k(a_j)}
{T_j\tau(\bbox{n})\over T_i\tau(\bbox{n})}.
$$
It is easy to check that all the constraints are satisfied
by the $\tau$-functional substitution. Moreover,
if we use this substitution for Darboux equations
(\ref{dDarboux}), what we get is just a standard
Hirota bilinear difference equation
$$
a_j(a_k-a_i)(T_iT_k \tau)T_j\tau+
a_k(a_i-a_j)(T_iT_j\tau)T_k\tau+
a_i(a_j-a_k)(T_jT_k\tau)T_i\tau=0.
$$
{\bf Remark 1}
The existence of constraints (\ref{constraints}) 
is not limited by the case of one-component KP
hierarchy and discrete times labeled by the
continuous parameter ($a_i$); in fact what we really need
to get these constraints is common zero of the functions
$g_i(\lambda)$ corresponding to discrete shifts $T_i$
(see \cite{AB2} for more details).
In the case when all three discrete variables
are essentially discrete, corresponding to four-component
KP hierarchy (there are three independent essentially discrete 
variables in the four-component case)
the $\tau$-functional substitution resolving the constraints
looks like
$$
\beta_{ij}(\bbox{n})=
{T_jT_i^{-1}\tau(\bbox{n})\over\tau(\bbox{n})},
$$
and using it, we get from (\ref{dDarboux})
an addition formula for the four-component
KP case containing only essentially discrete shifts
$$
(T_iT_k \tau)T_j\tau+
(T_iT_j\tau)T_k\tau+
(T_jT_k\tau)T_i\tau=0.
$$
Intermediate cases (i.e. when some of the discrete
variables are essentially discrete, while others
contain continuous parameter) are also possible.\\
{\bf Remark 2}
It is interesting to note that the set of relations
(\ref{constraints}) arises also in projective
geometry in the description of coordinate systems
in the incidence space (\cite{Hodge}, p. 260-262).

There is an important question what kind of system of
planar quadrilateral lattices are described
by this special case (reduction) of the discrete Darboux system
(corresponding to the one-component KP case).
To understand it, we should recover that the
function $\Phi$ satisfying the equation (\ref{1KPSM})
is connected with the {\em radius-vector} of
the related system of lattices.
More explicitly,
$$
\bbox{r}(\bbox{n})=(\Phi_1,\Phi_2,\Phi_3),
$$
where $\Phi_1,\Phi_2,\Phi_3$ are solutions to (\ref{1KPSM})
obtained from the same CBA function integrated with different
weight functions.

So geometrical characterization of
this special system of planar quadrilateral discrete
surfaces connected with the one-component KP hierarchy
is the following:
it is the three-dimensional
lattice built
of planar quadrilaterals
such that the projection of the radius-vector
to each coordinate axis satisfies equation (\ref{1KPSM}).

The case of equations (\ref{dDarboux}) + constrains
(\ref{constraints})
corresponding to the multicomponent
KP hierarchy will lead us to the matrix equation
(\ref{KPsingman}), where the matrix contains several
radius-vectors as columns, and we do not have clear
geometric interpretation for this case yet.

There is another geometric interpretation
for equation (\ref{1KPSM}) if the function
$\Phi$ is complex-valued. Then this equation
defines two conditions for the
system of hexagons
in the complex plane
with the vertices $$(T_1\Phi, T_1T_3\Phi,T_3\Phi,
T_3T_2\Phi, T_2\Phi, T_1T_2\Phi).$$
These hexagons are not necessarily convex and may even have
self-intersections.
The first condition is that the sum of the angles
at the vertices $T_1\Phi,T_3\Phi,
T_2\Phi$  (and also at the vertices
$T_1T_3\Phi,
T_3T_2\Phi,T_1T_2\Phi)$ is $2\pi n$.
The second is that the products
of lengths of three non-adjacent sides
are equal
$$
|T_1\Phi, T_1T_3\Phi||T_3\Phi,
T_3T_2\Phi||T_2\Phi, T_1T_2\Phi|=
|T_1T_3\Phi,T_3\Phi||T_3T_2\Phi, T_2\Phi||T_1T_2\Phi T_1\Phi|.
$$
Similar interpretation also exists for equation
(\ref{TodaSM}).
\subsection*{Acknowledgments}
The first author (LB) is grateful to the
Dipartimento di Fisica dell'Universit\`a
and Sezione INFN, Lecce, for hospitality and support;
(LB) also acknowledges partial support from the
Russian Foundation for Basic Research under grants
No 98-01-00525 and 96-15-96093 (scientific schools).


\end{document}